\documentclass[twocolumn,printnumbers,amsmath,amssymb,showpacs,prl]{revtex4}
\usepackage{graphicx}
\usepackage{color}

\begin{document}
\title{Finite size analysis of zero-temperature jamming transition under applied shear stress}

\date{\today}

\author{Hao Liu}
\author{Xiaoyi Xie$^{\dagger}$}
\author{Ning Xu$^*$}

\affiliation{CAS Key Laboratory of Soft Matter Chemistry, Hefei National Laboratory for Physical Sciences at the Microscale, and Department of Physics, University of Science and Technology of China, Hefei 230026, People's Republic of China.}

\begin{abstract}
By finding local minima of an enthalpy-like energy, we can generate jammed packings of frictionless spheres under constant shear stress $\sigma$ and obtain the yield stress $\sigma_y$ by sampling the potential energy landscape.  For three-dimensional systems with harmonic repulsion, $\sigma_y$ satisfies the finite size scaling with the limiting scaling relation $\sigma_y\sim\phi - \phi_{_{c,\infty}}$, where $\phi_{_{c,\infty}}$ is the critical volume fraction of the jamming transition at $\sigma=0$ in the thermodynamic limit.  The width or uncertainty of the yield stress decreases with decreasing $\phi$ and decays to zero in the thermodynamic limit.  The finite size scaling implies a length $\xi\sim (\phi-\phi_{_{c,\infty}})^{-\nu}$ with $\nu=0.81\pm 0.05$, which turns out to be a robust and universal length scale exhibited as well in the finite size scaling of multiple quantities measured without shear and independent of particle interaction.  Moreover, comparison between our new approach and quasi-static shear reveals that quasi-static shear tends to explore low-energy states.
\end{abstract}


\maketitle

At zero temperature and shear stress, a packing of frictionless spheres interacting via repulsions jams into a disordered solid when its volume fraction $\phi$ exceeds a critical value $\phi_c$ at the so-called Point J \cite{liu,ohern,van_hecke}.  As a simplified model to understand the noncrystalline liquid-solid transition of various materials including granular materials, foams, colloids, emulsions, and glasses, jammed packings of frictionless spheres exhibit interesting but unusual critical behaviors at Point J \cite{ohern,durian,silbert1,wyart1,ellenbroek,olsson,tighe,hatano,goodrich,xu1,zhao,drocco,ozawa,keys,head}.

In addition to the volume fraction, shear stress $\sigma$ and temperature $T$ have been proposed as the other two control parameters to cause generalized jamming transition, {\it i.e.}~yielding and glass transition \cite{liu}.  A jammed solid remains rigid when subject to a shear stress smaller than the yield stress $\sigma_y$, while it unjams and flows otherwise.  It has been shown that the yield stress of the $T=0$ jammed solids decreases with decreasing the volume fraction and vanishes at Point J \cite{olsson,tighe,pica_ciamarra,heussinger}.  This is different from the glass transition temperature at which a supercooled liquid is supposed to freeze into a glass, through the fact that in the $T=0$ limit glass transition occurs at a volume fraction lower than $\phi_c$ \cite{ikeda,krzakala,berthier2,parisi,zhang,wang,olsson1}.  Therefore, Point J is more relevant to the volume fraction and shear stress than to the temperature.  Jammed packings of frictionless spheres under applied shear stress thus serve as typical systems to study the criticality of Point J \cite{olsson,tighe,hatano}.

In most of the previous simulations, the yield stress of a jammed solid has been defined as either the average shear stress of the quasi-static shear flow in which the shear stress is not a controllable parameter \cite{berthier,pica_ciamarra,heussinger,xu2} or the critical shear stress extrapolated from nonequilibrium molecular dynamics simulations above which the system loses shear rigidity and flows forever \cite{xu2,pica_ciamarra}.  In the potential energy landscape perspective, the yield stress corresponds to the critical shear stress above which there is no jammed state which can sustain the applied shear stress.  Or in practice, the probability of finding such jammed states is low.  If states constrained at desired shear stress were quickly generated, we would be able to sample the potential energy landscape and locate the jamming transition at $\sigma>0$, in similar way to what was done for the jamming transition at $\sigma=0$ \cite{ohern}.  However, such an approach is apparently lacking.

In this letter, we report that the sampling of the potential energy landscape under the constraint of constant shear stress can be realized by looking for jammed states via minimization of an enthalpy-like energy.  The yield stress determined from the probability of finding jammed states is scaled well with the volume fraction.  The finite size scaling of the yield stress indicates that Point J is a critical point associated with a diverging length.  We find the same length scale in the finite size scaling of multiple quantities under zero shear stress and with different particle interactions, implying the universality of the length scale.  The width or uncertainty of the yield stress is inversely proportional to the square root of the system size, indicating that there is a well-defined yield stress in the thermodynamic limit.  Moreover, by comparing properties of jammed states obtained from our new approach and quasi-static shear, we find that quasi-static shear explores low-energy states in the potential energy landscape, which may provide us with a possible way to search for ultrastable glasses.

Our systems are three-dimensional with side length $L$ in all directions.  Lees-Edwards boundary conditions are applied to mimic shearing \cite{allen}.  To avoid crystallization, we put $N/2$ large and $N/2$ small spheres with equal mass $m$ in the system.  The diameter ratio of the large to small particles is $1.4$.  The interaction potential between particles $i$ and $j$ is
\begin{equation}
U_{ij}=\frac{\epsilon}{\alpha}\left( 1-\frac{r_{ij}}{d_{ij}}\right)^{\alpha}\Theta\left(1-\frac{r_{ij}}{d_{ij}}\right),\label{potential}
\end{equation}
where $r_{ij}$ is their separation, $d_{ij}$ is the sum of their radii, and $\Theta(x)$ is the Heaviside function.  Here we only show results for harmonic repulsion with $\alpha=2$.  To obtain jammed states at desired shear stress $\sigma$, we start with random high-temperature states and minimize the enthalpy-like energy
\begin{equation}
H(\vec{r}_1,...,\vec{r}_N,\gamma)=U(\vec{r}_1,...,\vec{r}_N,\gamma) - \sigma\gamma L^3,
\end{equation}
using FIRE minimization method \cite{fire}, where $U=\sum_{i=1}^{N-1}\sum_{j=i+1}^N U_{ij}$ is the internal energy, $\gamma$ is the shear strain, and $\vec{r}_i$ is the location of particle $i$.  We set the units of mass, energy, and length to be $m$, $\epsilon$, and small particle diameter $d_s$.

\begin{figure}
\includegraphics[width=0.48\textwidth]{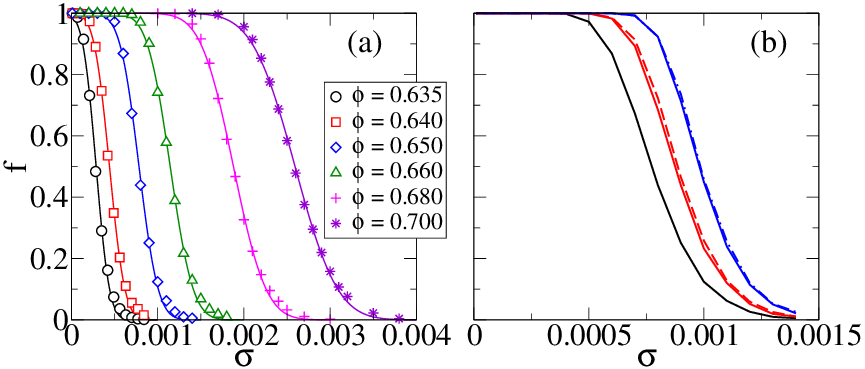}
\caption{\label{fig:fig1} (color online).  (a) Probability of finding jammed states, $f$, measured as a function of applied shear stress $\sigma$ with the cutoff strain $\gamma_c=1$ for $N=64$ systems.  The solid curves are the fits using Eq.~(\ref{erf}).  (b) Probability measured at $\phi=0.650$ with $\gamma_c=1$, $2$, and $4$ (solid curves from the left to the right).  The dashed and dot-dashed curves are predicted $\gamma_c=2$ and $\gamma_c=4$ curves using Eq.~(\ref{recur}).}
\end{figure}

The shear strain is initially set to be zero.  When the applied shear stress is much smaller than the yield stress, our algorithm can quickly find jammed states at small $\gamma$.  With increasing the shear stress, larger $\gamma$ is needed.  When $\sigma>\sigma_y$, because there is no static state being able to sustain the shear stress, $\gamma$ goes to infinity.  It is impracticable to run simulations up to extremely large $\gamma$.  We thus set a cutoff strain $\gamma_c$.  If the simulation does not find any local energy minima before $\gamma_c$, the initial random configuration corresponds to an unjammed state.  Otherwise, it leads to a jammed state.  For each pair of $\phi$ and $\sigma$, we enumerate the number of jammed states over $10000$ independent runs, from which the probability of finding jammed states, $f(\sigma,\phi)$, is determined.

Figure~\ref{fig:fig1}(a) shows $f(\sigma,\phi)$ with $\gamma_c=1$ for $N=64$ systems.  At fixed $\phi$, we approximate $f(\sigma,\phi)$ into a complementary error function
\begin{equation}
f(\sigma,\phi)=\frac{1}{2}{\rm erfc}\left[ \frac{\sigma - \sigma_y(\phi)}{\sqrt{2}w(\phi)}\right],\label{erf}
\end{equation}
where $\sigma_y$ and $w$ are the mean value [$f(\sigma_y,\phi)=0.5$] and standard deviation, which are defined here as the yield stress and its width.

With Lees-Edwards boundary conditions, the shear strain has a period of one.  A state under a strain of $\gamma+l$ does not make any difference from that under a strain of $\gamma$, where $l$ is an integer.  In this sense, $\gamma_c=1$ is a natural and reasonable choice.  Of course, $\sigma_y$ and $w$ vary with $\gamma_c$.  However, this variation is trivial and can be predicted simply from the recurrence relation
\begin{equation}
f_{l+1}=f_l+f(1-f_l),\label{recur}
\end{equation}
assuming that from $\gamma=l$ to $\gamma=l+1$ those $1-f_l$ unjammed states have another chance to go to jammed states with a probability of $f$,
where $f$, $f_l$, and $f_{l+1}$ are probabilites of finding jammed states at the same shear stress with $\gamma_c=1$, $l$, and $l+1$.  In Fig.~\ref{fig:fig1}(b), we compare $f_l$ obtained from the direct measure by setting $\gamma_c=l$ and predicted from Eq.~(\ref{recur}).  They agree well.  Therefore, the probability of finding jammed states measured within a unit strain, $f$, reflects nontrivially the fraction of configurational space occupied by jammed states under the constraint of constant shear stress.

\begin{figure}
\includegraphics[width=0.48\textwidth]{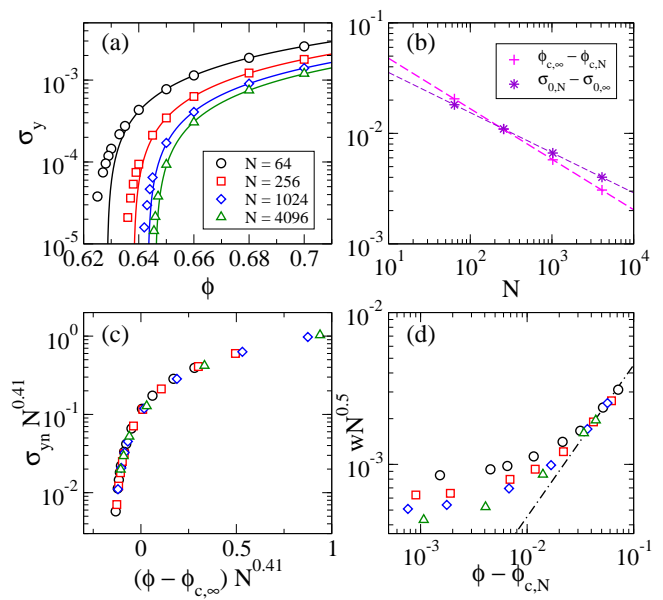}
\caption{\label{fig:fig2} (color online).  (a) Volume fraction $\phi$ and system size $N$ dependence of the yield stress $\sigma_y$.  The solid curves are the fits to the high $\phi$ data using Eq.~(\ref{yield_scaling}).  (b) System size dependence of the fitting parameters $\phi_{_{c,N}}$ and $\sigma_{_{c,N}}$ in Eq.~(\ref{yield_scaling}).  The dashed lines are the fits using Eqs.~(\ref{phic}) and (\ref{sigma0}).  (c) Finite size scaling of the reduced yield stress $\sigma_{yn}$.  (d) Volume fraction dependence of the width of the yield stress $w$ multiplied by $\sqrt{N}$.  The dot-dashed line shows the scaling of Eq.~(\ref{width}).}
\end{figure}

Figure~\ref{fig:fig2}(a) shows the volume fraction dependence of the yield stress measured for different system sizes.  At high volume fraction regime studied here, {\it e.g.}~from $\phi=0.65$ to $0.70$, the yield stress is linearly scaled with the volume fraction:
\begin{equation}
\sigma_y(\phi, N) = \sigma_{_{0,N}}(\phi - \phi_{_{c,N}}),\label{yield_scaling}
\end{equation}
where $\sigma_{_{0,N}}$ and $\phi_{_{c,N}}$ are fitting parameters.  As shown in Fig.~\ref{fig:fig2}(b), $\sigma_{_{0,N}}$ and $\phi_{_{c,N}}$ can be well fitted by the following scaling relations:
\begin{eqnarray}
\phi_{_{c,N}} &=& \phi_{_{c,\infty}} - 0.137N^{-0.457}, \label{phic}\\
\sigma_{_{0,N}} &=& \sigma_{_{0,\infty}} + 0.082N^{-0.363}, \label{sigma0}
\end{eqnarray}
where $\phi_{_{c,\infty}}=0.649$ and $\sigma_{_{0,\infty}}=0.018$ are the critical volume fraction of Point J and the slope of $\sigma_y(\phi)$ in the thermodynamic limit.  The value of $\phi_{_{c,\infty}}$ is consistent with previous studies of the same bi-disperse systems \cite{ozawa,wang}.  For finite size systems, $\sigma_y(\phi, N)$ deviates from Eq.~(\ref{yield_scaling}) at low volume fractions.  As illustrated by Fig.~\ref{fig:fig2}(a), this deviation is weaker with increasing the system size, which is likely associated with the system size dependence of the $\phi_c$ distribution \cite{ohern,xu3}.

The critical scaling of Eq.~(\ref{yield_scaling}) inspires us to perform finite size scaling of the yield stress.  As shown in Fig.~\ref{fig:fig2}(c), the yield stress indeed exhibits excellent finite size scaling, again suggesting that Point J at $\phi_{_{c,\infty}}$ is critical.  All the yield stress data collapse nicely onto a master curve in the following form:
\begin{equation}
\sigma_{yn} = \frac{\sigma_y}{\sigma_{_{0,N}}} = (\phi - \phi_{_{c,\infty}}) g_{_{\sigma}}\left[(\phi - \phi_{_{c,\infty}}) N^{\mu}\right], \label{yield_finite}
\end{equation}
with the limiting scaling of the yield stress being $\sigma_y\sim \phi - \phi_{_{c,\infty}}$, where $\mu=0.41\pm0.02$ is obtained to best collapse all the data.  Equation~(\ref{yield_finite}) implies a length $\xi$ diverging at $\phi_{_{c,\infty}}$ in the form:
\begin{equation}
\xi\sim (\phi - \phi_{_{c,\infty}})^{-\nu}, \label{length}
\end{equation}
where $\nu=\frac{1}{3\mu}=0.81\pm0.05$.
Later we will show that this length is not limited to the yield stress.  Finite size scaling of multiple quantities measured without shear exhibit the same length scale.

In previous studies \cite{pica_ciamarra,xu2}, two yield stresses have been observed from the liquid and solid perspectives, respectively.  The width of the yield stress $w$ directly reflects this uncertainty.  Figure~\ref{fig:fig2}(d) shows that $w$ decreases with decreasing the volume fraction, consistent with the observation of the decay of the gap between two yield stresses approaching Point J \cite{pica_ciamarra}.  In the high volume fraction regime studied here, $w$ is inversely proportional to $\sqrt{N}$.  The decay of $w$ at low volume fractions is even faster than $1/\sqrt{N}$.  With increasing the system size, $w(\phi, N)$ shows the tendency to approach the form:
\begin{equation}
w(\phi, N)\sim \frac{1}{\sqrt{N}}(\phi - \phi_{_{c,N}}) \label{width}.
\end{equation}
We thus claim that in the thermodynamic limit the yield stress is well-defined with $w=0$.  The gap between two yield stresses should be finite size effect, as already suggested in Ref.~\cite{xu2}.

\begin{figure}
\includegraphics[width=0.48\textwidth]{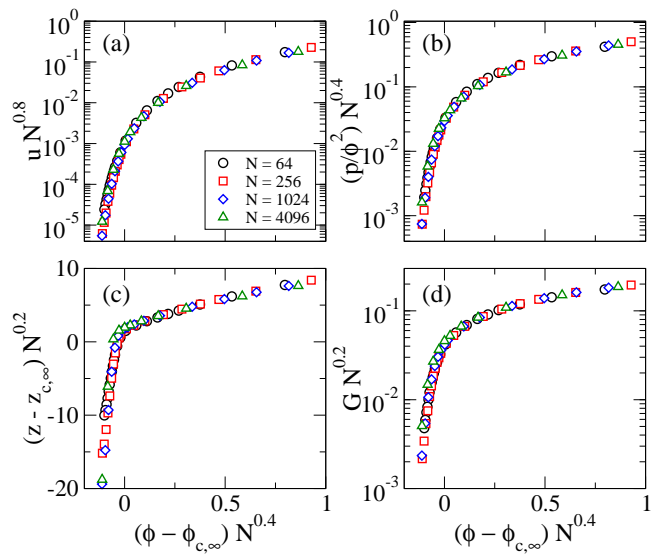}
\caption{\label{fig:fig3} (color online).  Finite size scaling of (a) potential energy per particle $u$, (b) pressure $p$, (c) excess coordination number beyond isostaticity $z-z_{_{c,\infty}}$, and (d) shear modulus $G$ for systems without shear.}
\end{figure}

To check whether the length described by Eq.~(\ref{length}) is universal or just specific to the yield stress, we perform the finite size analysis of typical quantities mostly concerned in the study of jamming under zero shear stress.  For each pair of $\phi$ and $N$, we generate $10000$ independent states without shear at $T=0$ (including jammed and unjammed) and then do the average.  Figure~\ref{fig:fig3} demonstrates that the potential energy per particle $u=U/N$, pressure $p$, coordination number $z$, and shear modulus $G$ all show very nice finite size scaling with the same length scale as proposed by Eq.~(\ref{length}):
\begin{eqnarray}
u&=&(\phi-\phi_{_{c,\infty}})^2g_{_u}\left[ (\phi-\phi_{_{c,\infty}}) N^{0.4} \right], \label{U}\\
\frac{p}{\phi^2}&=&(\phi-\phi_{_{c,\infty}})g_{_p}\left[ (\phi-\phi_{_{c,\infty}}) N^{0.4} \right], \label{p}\\
z-z_{_{c,\infty}}&=&(\phi-\phi_{_{c,\infty}})^{0.5}g_{_z}\left[(\phi-\phi_{_{c,\infty}}) N^{0.4} \right],\label{z}\\
G&=&(\phi-\phi_{_{c,\infty}})^{0.5}g_{_G}\left[ (\phi-\phi_{_{c,\infty}}) N^{0.4}\right], \label{G}
\end{eqnarray}
where $z_{_{c,\infty}}=6$ is the isostatic value.  We tune the exponent of $N$ from $0.41$ to $0.4$ to have the best data collapse.  The limiting scaling relations of these quantities are well known for marginally jammed solids with harmonic repulsion \cite{ohern,durian,zhao,wang}.  Equations~({\ref{U}}) and (\ref{p}) are simply related by the relation $p\sim\phi^2\frac{{\rm d}U}{{\rm d}\phi}$, which leads to the factor $\phi^2$ in Eq.~(\ref{p}) and the relation $g_p(x)\sim2g_u(x)+xg_u'(x)$.  The same length scale from the finite size scaling of the yield stress, potential energy, pressure, coordination number, and shear modulus suggests that the scaling exponent of the length found here is universal for three-dimensional jammed states with harmonic repulsion.

It has been shown that most of the scaling relations of marginally jammed states depend on the exponent $\alpha$ in Eq.~(\ref{potential}) of the particle interaction, except for the coordination number \cite{ohern}.  It is thus possible that the finite size scaling for the coordination number shown in Fig.~\ref{fig:fig3}(c) also works for other inter-particle potentials like Hertzian repulsion [$\alpha=2.5$ in Eq.~(\ref{potential})]. If then, the same length scale would be observed in the finite size scaling of multiple quantities for Hertzian repulsion as well.  We repeat Fig.~\ref{fig:fig3} for Hertzian repulsion (not shown here) and indeed find the same length scale.  Therefore, the length scale reported here may be independent of inter-particle potential, at least for harmonic and Hertzian repulsions.

\begin{figure}
\includegraphics[width=0.48\textwidth]{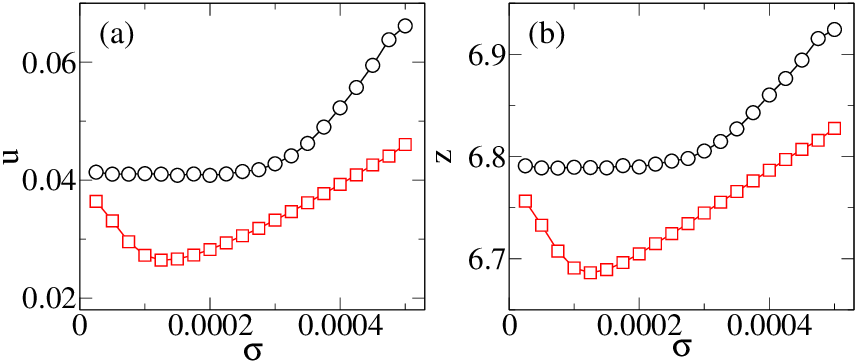}
\caption{\label{fig:fig4} (color online).  Shear stress $\sigma$ dependence of (a) the potential energy per particle $u$ and (b) coordination number $z$ obtained from random sampling (circles) and quasi-static shear sampling (squares).  The systems consist of $N=1024$ particles at $\phi=0.66$.  The solid curves are to guide the eye. }
\end{figure}

Minimizing the enthalpy-like potential enables us to sample jammed states under desired shear stress.  Because the initial random states before minimization are independently selected, our sampling of the potential energy landscape is unbiased.  We can thus have an unbiased statistical picture about how the properties of jammed solids vary with shear stress.   In Fig.~\ref{fig:fig4}, we show the potential energy per particle and coordination number averaged over jammed states under the same shear stress obtained from the unbiased random sampling.  With increasing the shear stress, both the potential energy and coordination number show a plateau at low shear stresses and shoot up near yielding.

As a comparison, we also show the results for quasi-static shear sampling in Fig.~\ref{fig:fig4}.  To mimic quasi-static shear, we successively deform jammed states from $\gamma=0$ to $1$ using a step strain $\Delta\gamma=10^{-4}$ followed by the potential energy minimization.  $10000$ jammed states with different shear stresses are obtained during one course of quasi-static shear, from which the shear stress dependence of the potential energy and coordination number can be achieved.  The results in Fig.~\ref{fig:fig4} are from $1000$ independent runs of quasi-static shear.  In contrast to our random sampling, quasi-static shear leads to a decrease of both the potential energy and coordination number with increasing the shear stress at low shear stresses.  At all shear stresses, jammed states found by quasi-static shear sampling always have lower potential energy and coordination number than those obtained from random sampling.  This discrepancy implies the biased nature of the quasi-static shear to sample the potential energy landscape: it tends to explore low-energy states.

The bias of quasi-static shear sampling contains some interesting implications.  For jammed states interacting via repulsions, lower coordination number and potential energy mean that the states are closer to the unjamming transition subject to the change of volume fraction.  This explains why the critical volume fraction of the jamming transition determined from quasi-static shear sampling is higher than that from random sampling at $\sigma=0$ \cite{heussinger,olsson}.

More interestingly, for the case shown in Fig.~\ref{fig:fig4}, quasi-static shear can find states with potential energy about $35\%$ lower than the normal value.  This is actually analogous to the fact that inherent structures with lower potential energy can be explored by supercooled liquids with slower cooling rate \cite{sastry}.  However, the advantage of quasi-static shear is that it can overcome energy barriers easily and thus efficiently speed up the search of low-energy states.  Recently, it has been reported that ultrastable glasses with low energy and aged over thousands of years can be quickly obtained from vapor deposition method \cite{swallen,singh}.  Since quasi-static shear tends to explore low-energy states, it may provide us with an alternate efficient way to search for ultrastable glasses.

In conclusion, we sample the potential energy landscape under the constraint of constant shear stress by minimizing the enthalpy-like energy.  Using this new approach, we obtain the yield stress of jammed solids from measuring the probability of finding jammed states under constant shear stress.  The yield stress and multiple quantities measured without shear all show very nice finite size scaling, from which we obtain a universal length scale described in Eq.~(\ref{yield_finite}).  Multiple length scales have been reported for the jamming transition at Point J in different measurements \cite{ohern,silbert1,wyart1,ellenbroek,olsson,hatano,goodrich,drocco,ozawa}, which may also be one of the most special and elusive features of the criticality of Point J. The length scale reported here is robust for three-dimensional systems because it is associated with multiple quantities and possibly independent of the inter-particle potential.  Moreover, we propose to look for low-energy ultrastable glasses using quasi-static shear because it can explore low-energy states efficiently.

We are grateful to Kunimasa Miyazaki and Stephen Teitel for helpful discussions.  This work is supported by National Natural Science Foundation of China No. 11074228 and 91027001, National Basic Research Program of China (973 Program) No. 2012CB821500, CAS 100-Talent Program No. 2030020004, and Fundamental Research Funds for the Central Universities No. 2340000034.

\end{document}